\documentclass[aps,prd,showpacs,preprintnumbers]{revtex4}
\usepackage{graphics}
\def \beq{\begin{equation}}
\def \eeq{\end{equation}}
\def \beqa{\begin{eqnarray}}
\def \eeqa{\end{eqnarray}}

\def\ie{{\sl i.\ e.\/}}
\def \ipr{P^{\mathbf 1}_2}
\def \iprc{P^{\mathbf *}_2}

\def \tr{{\rm Tr\,\/}}

\def\etal{{\sl et al.\/}}

\def\npb{{\sl Nucl.\ Phys.\/}, B}
\def\plb{{\sl Phys.\ Lett.\/}, B}
\def\prb{{\sl Phys.\ Rev.\/}, B}
\def\prd{{\sl Phys.\ Rev.\/}, D}
\begin{document}
\title{Eigenvalues and Eigenvectors of the Staggered Dirac Operator
   at Finite Temperature}
\author{R.\ V.\ \surname{Gavai}}
\email{gavai@tifr.res.in}
\affiliation{Department of Theoretical Physics, Tata Institute of Fundamental
         Research,\\ Homi Bhabha Road, Mumbai 400005, India.}
\author{Sourendu \surname{Gupta}}
\email{sgupta@tifr.res.in}
\affiliation{Department of Theoretical Physics, Tata Institute of Fundamental
         Research,\\ Homi Bhabha Road, Mumbai 400005, India.}
\author{R.\ \surname{Lacaze}}
\email{Robert.Lacaze@cea.fr}
\affiliation{Service de Physique Theorique, CEA Saclay,\\
         F-91191 Gif-sur-Yvette Cedex, France.}

\begin{abstract}
\end{abstract}

\begin{abstract}
We examine the eigenvalues and eigenvectors of the staggered Dirac
operator on thermal ensembles created in QCD with two flavours
of staggered quarks. We see that across the phase transition a gap opens
in the spectrum. For finite volume lattices in the low-temperature phase
the eigenvectors are extended, but generic field configurations in the
high temperature phase give rise to localized eigenstates. We examine
measures of the stability of such localization and find that at finite
volumes localization occurs through Mott's mechanism of the formation of
mobility edges. However, the band gap between the localized and extended
states seem to scale to zero in the limit of large volume.
\end{abstract}
\maketitle

\section{Introduction}

Any fermionic operator can be written in the spectral form
\beq
   \hat{\cal O} = \sum_{\lambda\mu}O_{\mu\lambda}|\lambda\rangle\langle\mu|,
\eeq
where $|\lambda\rangle$ is an eigenvector of the Dirac operator with
eigenvalue $\lambda$, evaluated separately on each configuration.
Typical operators of interest contain quark loops with various insertions,
\ie, ${\cal O}=\tr(A_1 (D+m)^{-1} A_2 (D+m)^{-1}\cdots A_n(D+m)^{-1})$. As a result,
\beq
   {\cal O} = \sum_{\lambda_1\cdots\lambda_n}
     \frac{\langle\lambda_1|A_1|\lambda_2\rangle
           \langle\lambda_2|A_2|\lambda_3\rangle\cdots
           \langle\lambda_n|A_2|\lambda_1\rangle}{\prod_{i=1}^n(m+\lambda_i)},
\eeq
where we use the symbol $D$ to refer to the massless Dirac operator.
If the $A_i$ commute with $D$, then the matrix elements in the
numerator are diagonal, and all questions about the operator reduce
to the simultaneous eigenvalues of the $A_i$ and the Dirac operator.
This happens, for example, in the chiral sector of the theory, where
one deals with questions about n-point functions of pions. Since
$\gamma_5 D\gamma_5 = D^\dag$, most questions about the chiral
sector can be answered if the eigenvalues are known. As a result,
the thrust of many previous studies of QCD to date has been on the
spectrum of eigenvalues, particularly on comparisons with random
matrix theory (RMT) \cite{rmt}. This focus is due to the fact that
RMT is known to be equivalent to chiral perturbation theory in some
limits \cite{equiv}.

However, at finite temperature, especially above $T_c$, chiral
perturbation theory is not the appropriate long-distance effective
theory. Furthermore, there are interesting questions at many different
length scales and one may need to build different effective theories
to answer these questions. Several questions involve fermionic loops
with insertions of operators which do not commute with $D$. 
An example is the vector susceptibility,
\beq
   \chi_V = \sum_{\lambda_1,\lambda_2}
     \frac{|\langle\lambda_1|\gamma_\mu|\lambda_2\rangle|^2}
                            {(m+\lambda_1)(m+\lambda_2)},
\label{vsus}\eeq
which includes quark number susceptibilities. Deeper understanding
of such quantities need the study of the eigenvectors \cite{modes}.

\section{Eigenvalues}

We analyzed configurations generated in the study of QCD with two
flavours of dynamical staggered quarks at a lattice spacing $a=1/4T$
\cite{endpt}.  The scale fixing yielded $T_c/m_\rho=0.186 \pm 0.006$.
As $T$ varied between $0.75T_c$ and $2T_c$, the renormalized quark
mass was kept constant.  The physical box size, $L=N_s a$ where
$N_s$ is the box size in units of the lattice spacing. The aspect
ratio was varied in the range $2\le LT\le6$.

We investigated the eigenvalues, $\lambda$, and eigenvectors,
$|\lambda\rangle$, of the massless Dirac operator, $D$, in typical thermal
ensembles picked from these simulations. We used five configurations
separated by two autocorrelation times at all temperatures and
volumes except at $1.05T_c$ where we verified the results using twenty
configurations.  Eigenvalues and eigenvectors were computed with the
ARPACK subroutines \cite{arpack}. For convergence, the tolerance is
chosen so that
\beq
   |r|^2 < \epsilon,\qquad{\rm where}\qquad
           r=(D-\lambda)\psi_\lambda,
\label{arpack}\eeq
where $\lambda$ is an eigenvalue and $\psi_\lambda$ is the corresponding
eigenvector. We report results with $\epsilon=2\times10^{-13}$.

\begin{figure}[htb]\begin{center}
   \rotatebox{270}{\scalebox{0.45}{\includegraphics{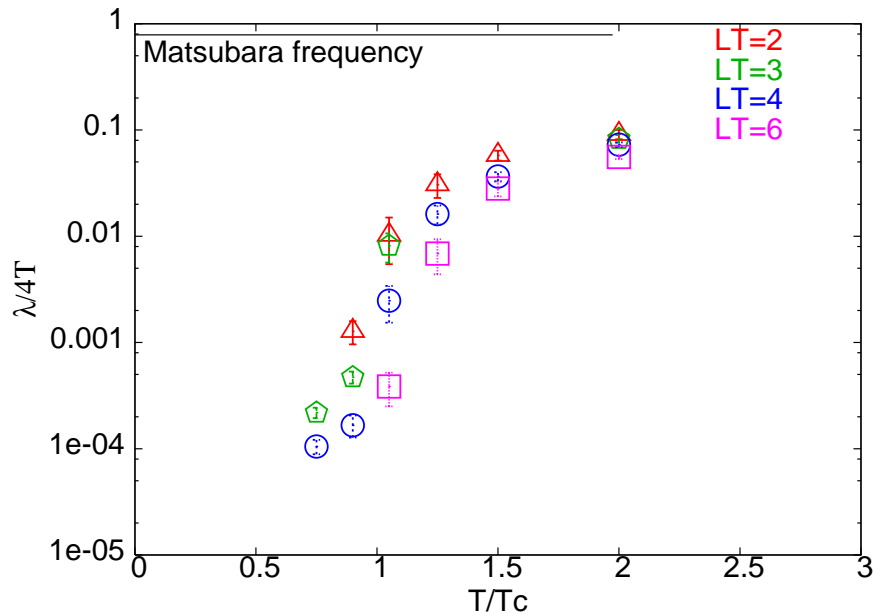}}}
   \end{center}
   \caption{The lowest Dirac eigenvalues as a function of $T$ for
    different spatial lattices. In all cases the lattice spacing
    $a=1/4T$. Also shown in the lowest Matsubara frequency expected
    at this cutoff, \ie, the expectation for free fermions.}
\label{fg.eigv}\end{figure}

\begin{table}[htb]
\begin{center}
\begin{tabular}{c|c|c|c|c|c|c|c|}
\hline 
& Lattice & cutoff & N points & $\chi^2/N$ & $a_1$ & $a_2$ & $a_3$ \\
\hline 
& $4\times 16^3$ & 0.01 &  189 & 0.77 & 0.153(3) & & \\
&                & 0.06 & 1147 & 0.37 & 0.153(2) & & \\
&                &      &      & 0.20 & 0.152(2) & 0.050(50) & \\
&                &      &      & 0.17 & 0.152(2) &  & 0.8(1.0) \\
&                &      &      & 0.15 & 0.153(2) & -0.08(5) & 1.9(1.0) \\
\hline
& $4\times 12^3$ & 0.02 & 153 & 1.00 & 0.147(2) & & \\
&                & 0.06 &  469 & 0.69 & 0.148(2) & & \\
&                & 0.14 & 1176 & 5.17 & 0.153(2) & & \\
&                &      &      & 0.27 & 0.144(2) & 0.116(26) & \\
&                &      &      & 0.20 & 0.147(2) & & 0.79(23) \\
&                &      &      & 0.16 & 0.146(2) & 0.046(26) & 0.49(23)\\
\hline
& $4\times 8^3$ & 0.06 & 128 & 0.17 & 0.135(11) & & \\
&               & 0.08 & 172 & 0.16 & 0.136(10) & & \\
&               & 0.10 & 222 & 0.23 & 0.138(10) & & \\
&               & 0.375 & 1150 & 6.80 & 0.175(7) & & \\
&               &       &      & 0.04 & 0.126(7) & 0.196(26) & \\
&               &       &      & 0.28 & 0.144(7) & & 0.43(8) \\
&               &       &      & 0.03 & 0.128(7) & 0.175(25) & 0.05(8) \\
\hline
\end{tabular}\end{center}
\caption{Fits of the cumulative density for $\beta=5.26$, \ie, $T/T_c=0.90\pm0.01$.}
\label{tb.lowt}\end{table}

\begin{table}[htb]
\begin{center}
\begin{tabular}{c|c|c|c|c|c|c|c|}
\hline
& Lattice & cutoff & N points & $\chi^2/N$ & $a_1$ & $a_2$ & $a_3$ \\
\hline 
& $4\times 24^3$ & 0.0025 &   33 & 0.20& 0.030(3)& & \\
&                & 0.01   & 140  & 0.22& 0.032(3)& & \\
&                & 0.031  & 535  & 0.04& 0.030(3)& 0.41(5)& \\
&                & 0.031  & 535  & 0.03& 0.031(3)& 0.25(10)& 5(2)\\
& $4\times 16^3$ & 0.01 &   22 & 0.38 & 0.0148(36) & & \\
&                & 0.06 &  323 & 5.35 & 0.0317(36) & & \\
&                &      &      & 1.89 &        & 0.81(9) & \\
&                &      &      & 0.03 & 0.0125(37) & 0.53(9) & \\
&                & 0.12 & 1077 & 0.03 & 0.0136(33) & 0.503(37) & \\
&                &      &      & 1.27 & 0.029(3) &  & 3.4(0.4) \\
&                &      &      & 0.01 & 0.0118(32) & 0.57(4) & -0.5(4) \\
\hline
& $4\times 12^3$ & 0.02 &  20 & 0.57 & 0.0135(45) & & \\
&                & 0.06 &  130 & 4.06 & 0.0291 & & \\
&                &      &      & 0.05 & 0.0060(46) & 0.62(11) & \\
&                & 0.20 & 1068 & 31.2 & 0.0767 & & \\
&                &      &      & 0.13 & 0.0147(39) & 0.45(3) & \\
&                &      &      & 2.18 & 0.039(4) & & 1.77(16) \\
&                &      &      & 0.01 & 0.008(4) & 0.581(26) & -0.55(16)\\
\hline
& $4\times 8^3$ & 0.01 &   9 & 0.09 & 0.047(25) & & \\
&               & 0.02 &  19 & 0.11 & 0.053(24) & & \\
&               & 0.06 &  68 & 0.16 & 0.064(18) & & \\
&               &      &     & 0.03 & 0.048(18) & 0.414& \\
&               & 0.20  &  373 & 0.01 & 0.052(11) & 0.348(26) & \\
&               & 0.415 & 1160 & 0.05 & 0.062(7) & 0.29(2) &    \\
&               &       &      & 0.003 & 0.051(7) & 0.38(2) & -0.16(6) \\
\hline
\end{tabular}\end{center}
\caption{Fits of the cumulative density for $\beta=5.30$, \ie, $T/T_c=1.05\pm0.01$.}
\label{tb.hight}\end{table}

The lowest staggered Dirac eigenvalue, $\lambda_0$, evolves 
as shown in Figure \ref{fg.eigv}.  There is
a clear crossover from low to high temperature behaviour evidenced by
an increase in the lowest eigenvalue by three order of magnitude in the
neighbourhood of $T_c$. This becomes sharper in the neighbourhood of $T_c$
with increasing spatial size of the lattice.

It is also worth noting that at $2T_c$ there seems to be little
remaining volume dependence. It is interesting that this large
volume behaviour sets in at a minimum $L$ given by $L\lambda_0
\approx 0.8$. Consistent with this observation, at $1.5T_c$ the
lattice sizes which satisfy this condition also give results which
are almost volume independent.

In Figure \ref{fg.eigv} we further show the lowest eigenvalue of the
free Dirac operator on a lattice of the same coarseness. In the limit of
zero lattice spacing this would correspond to the Matsubara frequency,
$\Omega=\pi T$. The full QCD configurations can be seen to lie very
far from the free field theory (ideal gas) limit at all temperatures up
to 2$T_c$.
This is consistent with another observation at the same lattice spacing
that the pseudoscalar screening correlator constructed with staggered
Dirac quarks yields a screening mass far below that expected from the
free theory \cite{mtc}.  It would be interesting to see whether this
correlation, related to the chiral behaviour, changes in the same way
due to various improvements in the gauge and fermion actions and in the
continuum limit.

The Banks-Casher formula \cite{banks} relates the density of Dirac
eigenvalues at zero with the chiral condensate. In order to utilize this
formula we expand the cumulative distribution of the eigenvalues in the
form \beq
   I(x) = \int_0^x d\lambda \rho(\lambda) = \sum_{n\ge1} a_n x^n,
   \qquad{\rm from\ which}\qquad \rho(x) = \sum_{n\ge1} na_n x^{n-1}.
\label{density}\eeq 
where the $\gamma_5$-Hermiticity of the staggered Dirac operator as well as 
its anti-Hermitean nature have been used.   Together they imply that
the eigenvalues are paired and imaginary, $\pm i\lambda$. The integration 
above is over the positive $\lambda$ values.  Note that the reflection 
symmetry in $\lambda$ permits the existence of an $a_2$ term (even $n$, in 
general) only if the $\rho(\lambda)$ is non-analytic at the origin.

The cumulative distribution was constructed numerically and fitted
to the form in eq.\ (\ref{density}) for configurations above and
below $T_c$. Indicative results are shown in Tables \ref{tb.lowt}
and \ref{tb.hight}.  The tables are arranged in increasing order of
$N$, the number of eigenvalues included, and $n$ of eq.(\ref{density}).
Note that below $T_c$ one gets a good determination of $a_1$ for cutoffs
of the order of 0.1 or so. In fact, the values of $a_1$ do not
depend on $n$ or $N$. With increasing lattice size, $L$, the estimate of $a_1$
increases marginally. On the larger lattices $a_2$ is compatible
with zero, indicating that the distribution is analytic.

Above $T_c$, $a_2$ is clearly non-zero for all cut-offs while $a_1$
drops with increasing $L$. This behaviour, shown in Table \ref{tb.hight},
implies that a non-analyticity develops in the spectral density. This
non-analyticity is due to the formation of a gap--- the spectral density
is exactly zero upto the gap, and then becomes non-zero.  Another way to
test this would be to introduce a gap explicitly in the eq.\ (\ref{density}),
\beq
   \rho(x) = \sum_{n\ge1} na_n (x-x_0)^{n-1}.
\label{gapped}\eeq
Indeed, when one does that, a non-zero value of the gap, $x_0$, is observed
for those temperatures where $a_2$ is non-zero by the other method.

\section{Eigenvectors and measures of localization}

\begin{figure}[htb]\begin{center}
   \rotatebox{270}{\scalebox{0.45}{\includegraphics{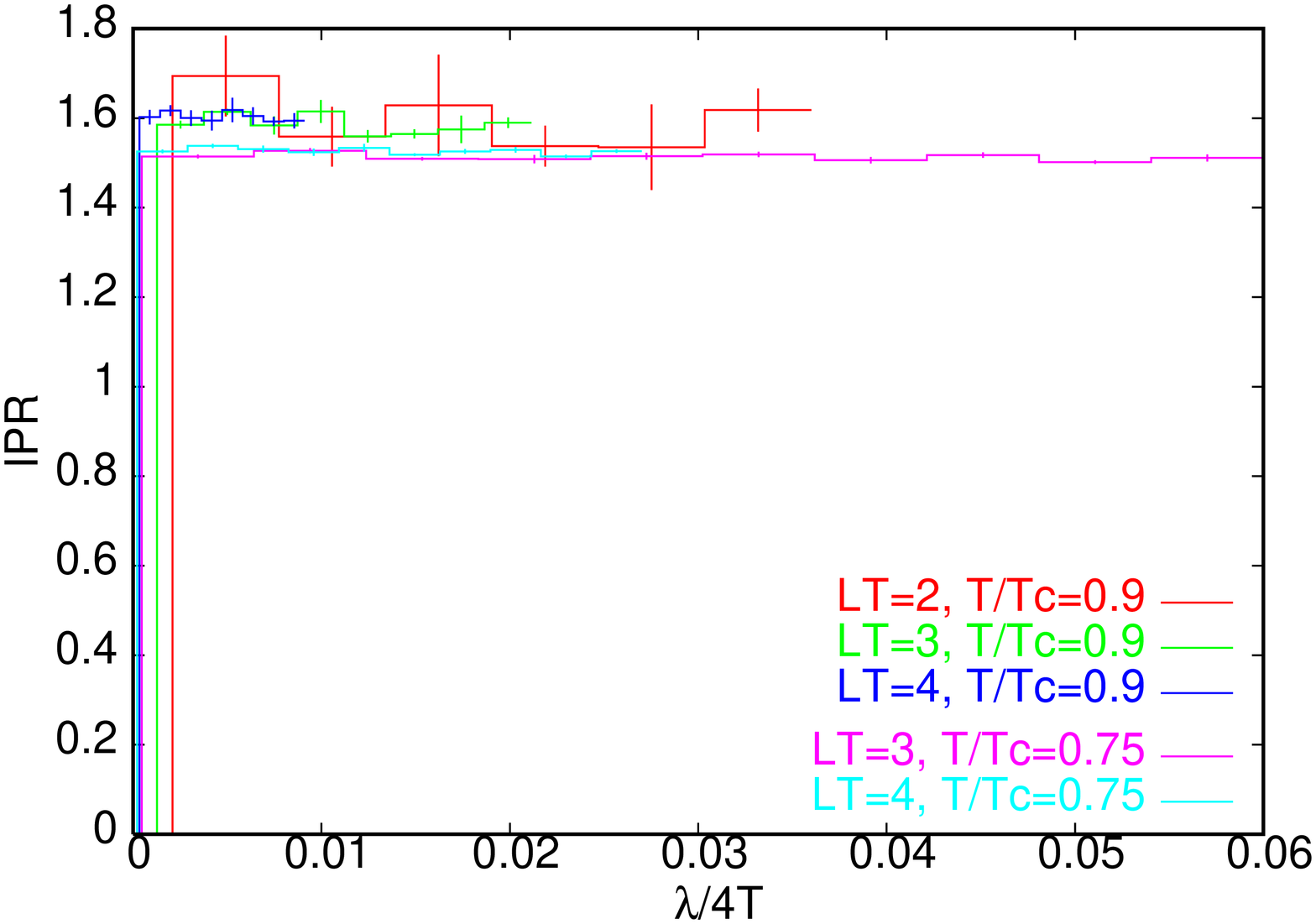}}}
   \rotatebox{270}{\scalebox{0.45}{\includegraphics{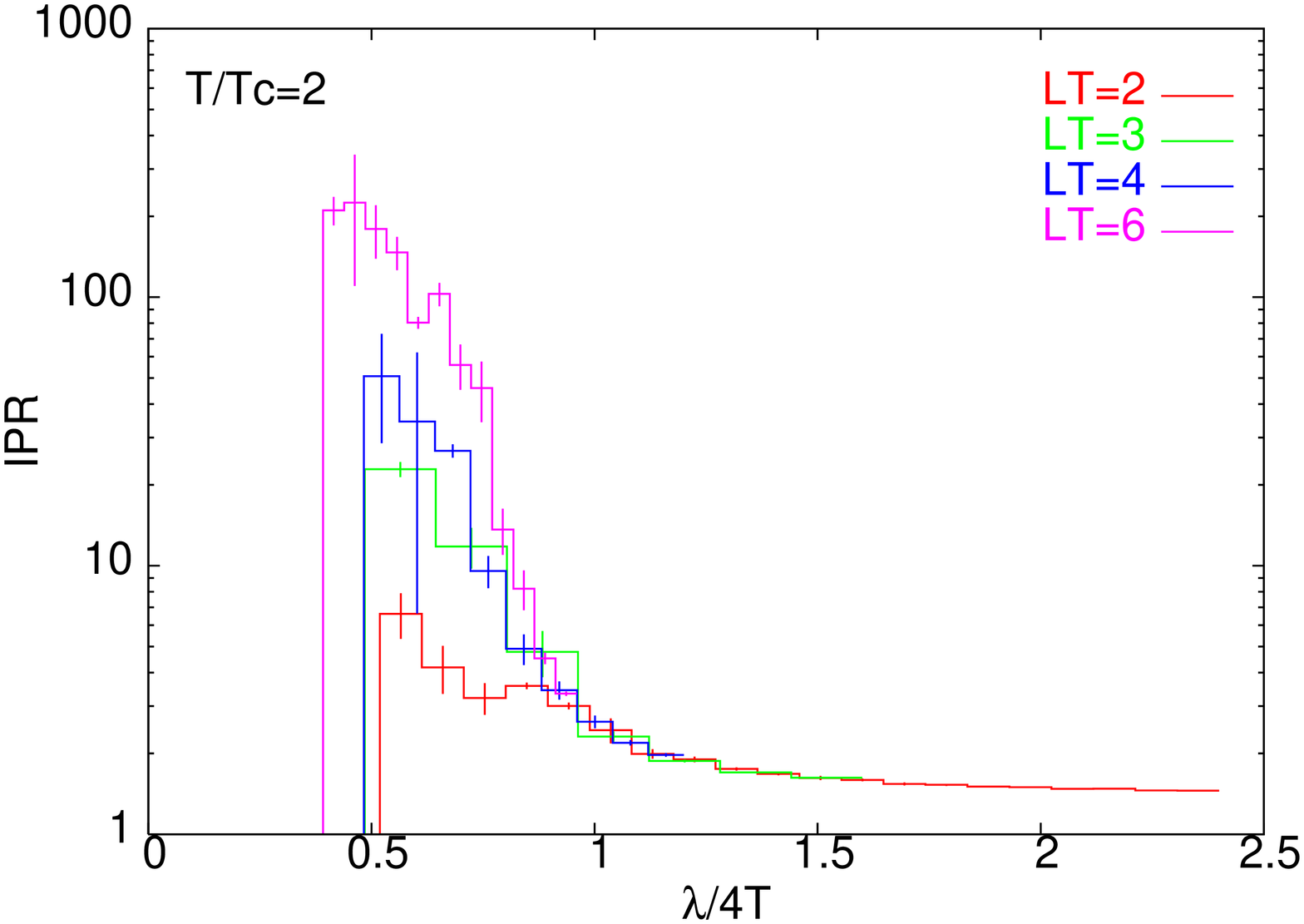}}}
   \end{center}
   \caption{The IPR, $\ipr$, above and below $T_c$ as a function
    of the staggered Dirac eigenvalue $\lambda$.}
\label{fg.ipr}\end{figure}

\begin{figure}[htb]\begin{center}
   \rotatebox{270}{\scalebox{0.45}{\includegraphics{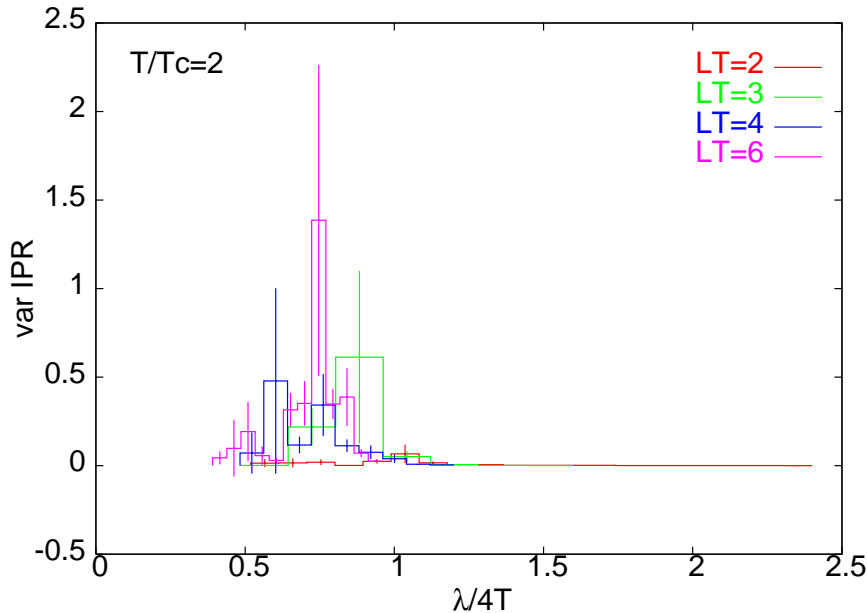}}}
   \end{center}
   \caption{The relative fluctuations in the IPR, \ie, the ratio of
    the variance and the mean, at $2T_c$ as a function of the staggered
    Dirac eigenvalue, $\lambda$.}
\label{fg.fluct}\end{figure}

\begin{figure}[htb]\begin{center}
   \scalebox{0.65}{\includegraphics{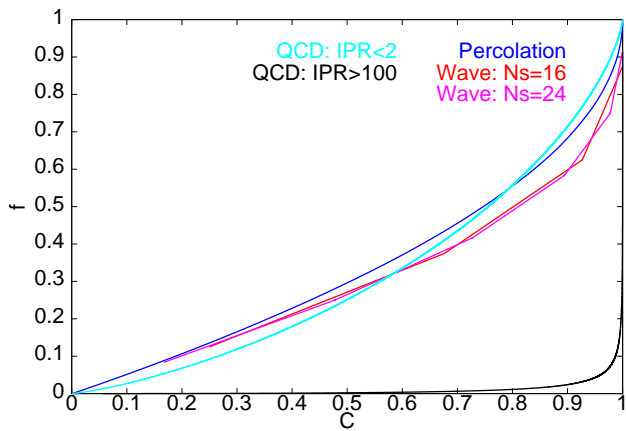}}
   \end{center}
   \caption{Examples of the localization function, $f({\cal C})$, for
    different models of localization, compared with two examples from
    QCD.}
\label{fg.horvath}\end{figure}

\begin{figure}[htb]\begin{center}
   \scalebox{0.65}{\includegraphics{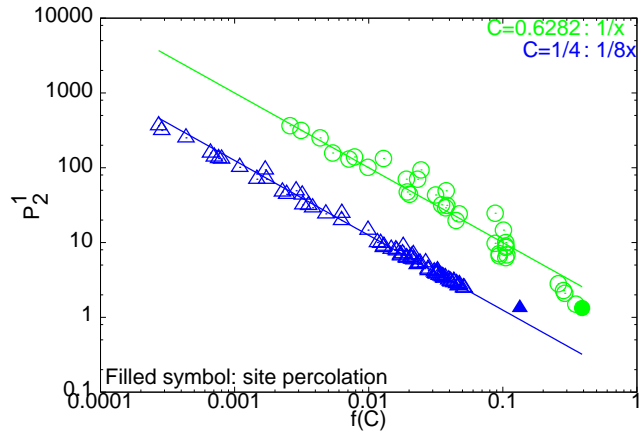}}
   \end{center}
   \caption{The strong correlation between IPR, $\ipr$, and the
    value of the localization function for two values of $\cal C$
    shows that the latter contains all the information available
    in the former.}
\label{fg.iprequiv}\end{figure}

\begin{figure}[htb]\begin{center}
   \scalebox{0.65}{\includegraphics{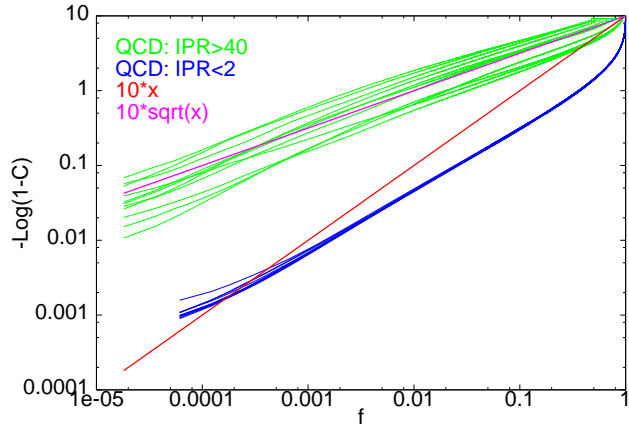}}
   \end{center}
   \caption{Scaling of the localization function in the vicinity of
    $f={\cal C}=1$ shows that staggered Dirac eigenvectors with
    IPR larger than 40 fall exponentially far from the peak, whereas
    those with IPR less than 2 have drastically different behaviour.}
\label{fg.expon}\end{figure}

\begin{figure}[htb]\begin{center}
   \scalebox{0.65}{\includegraphics{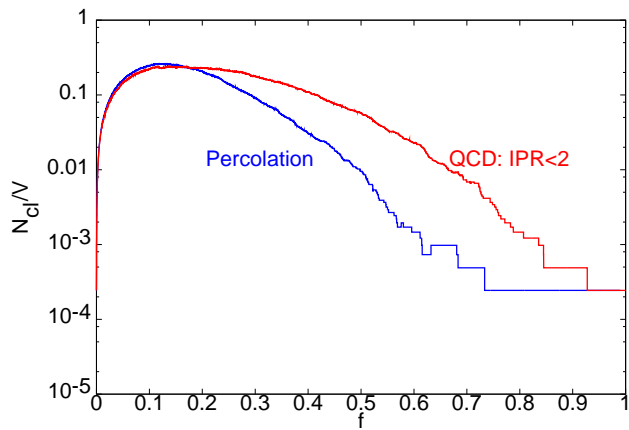}}
   \end{center}
   \caption{The number of clusters, $N_c$, normalized by the lattice
    volume for uncorrelated site percolation and for those eigenvectors
    of the staggered Dirac operator at $2T_c$ which have IPR smaller than 2.}
\label{fg.cluster}\end{figure}

The eigenvectors of the Dirac equation, $\psi$, are often investigated
through the localized moments
\beq
   P^\gamma_n(\lambda) = V^{n-1}\sum_r \left|p_\gamma(r;\lambda)\right|^n,
   \qquad{\rm where}\qquad p_\gamma(r;\lambda) = 
                      \langle\lambda|\gamma|\lambda\rangle,
   \qquad{\rm and}\qquad P^{\mathbf 1}_1=1,
\label{mom}\eeq
$\gamma$ is a matrix in Dirac space \footnote{Note that $P^V_2$ is
needed for the vector susceptibility defined in eq.\ (\ref{vsus})},
the inner product in the definition of $p_\gamma(r;\lambda)$ involves
a sum over spin-flavour and colour indices, the explicit sum is
over all $V$ lattice sites $r$, and the normalization of the
eigenvectors, $P^{\mathbf 1}_1$ involves the density where the Dirac
matrix is identity.  For staggered quarks one has the identity
$P^{\mathrm 1}_n = P^{\gamma_5}_n$. The second moment, $n=2$ is
called the inverse participation ratio (IPR). The moments $P^{\mathrm
1}_n$ have the interesting property that for constant $\psi=1/\sqrt
V$, one finds $P^{\mathrm 1}_n=1$, whereas for the localized
$\psi(r)=\delta_{r,r_0}$, one has $P^{\mathrm 1}_n = V^{n-1}$.

Histograms of IPR against $\lambda$ are shown in Figure \ref{fg.ipr}.
There is a very clear difference between the IPR observed below and
above $T_c$. Below $T_c$ the IPRs are close to unity, without any clear
dependence on the eigenvalues. In contrast, the situation is dramatically
different above $T_c$; several eigenvectors have very large values of
IPR. There is correlation between the eigenvalue and IPR, with larger
eigenvalues coming with substantially smaller IPR.

Below $T_c$ there is little sign of volume dependence of the IPR,
consistent with the small values seen there. Above $T_c$ the smaller
IPR values seen for large $\lambda$ are also volume independent.
However, as shown in Figure \ref{fg.ipr}, larger values of IPR are
volume dependent. A test of scaling shows that the lattice size
dependence is consistent with a power behaviour, $L^\alpha$, with
$2.5 \le \alpha \le 3.5$.  Again, this is not unexpected, since IPR
is constructed to be proportional to the volume for localized
eigenvectors.

In \cite{modes} the transition from volume dependent to independent
values of IPR is used to locate the ``mobility edge''. By this
identification one would have a mobility edge at $\lambda\simeq1.25T_c$
for a temperature of $2T_c$. However, the notion of a mobility edge
contains more physics and we shall examine it more critically in a
later section.

The eigenvalues and eigenvectors of the Dirac operator are clearly
dependent on the gauge field backgrounds. However, thermodynamic
quantities constructed from these have fluctuations which decrease
rapidly with increasing lattice size.  The IPR is not such a variable:
its fluctuations are comparable to the average, as can be seen in
Figure \ref{fg.fluct}. The ratio of the variance and mean of $\ipr$,
as a function of $\lambda$ at $2T_c$, is of order unity \footnote{In
condensed matter systems, such an observation would lead to a
prediction of conductance fluctuations in random media.}. The
localization properties of Dirac eigenfunctions can therefore serve
to classify the ensemble of gauge configurations which give important
contributions to the thermal path integral. This is an obvious
statement for overlap quarks, where localized chiral eigenvectors
of the overlap Dirac operator are closely connected to localized
gauge field configurations which are taken to be the lattice analogue
of instantons. It is interesting that localization using staggered
quarks, where the connection to topology is obscure, can also be
used as a tool for analysis of gauge configurations.

The notion of localization has been closely examined in \cite{horvath}.
Since $p_{\mathbf 1}(r)$ is non-negative and normalized to unity
one can construct a measure of localization in the following way.
Take a value $p_f$ and find the fraction of the lattice sites,
$f(p_f)$, containing values $p_{\mathbf 1}(r)>p_f$. Clearly $f(p_f)$
lies between 0 and 1, and is a decreasing function of $p_f$. The
integral of $p_{\mathbf 1}(r)$ over these sites, ${\cal C}(p_f)$,
lies between 0 and 1, and is another decreasing function of $p_f$.
Eliminating $p_f$ between these two, one obtains Horvath's localization
function $f({\cal C})$. Clearly $f({\cal C}=0)=0$, $f({\cal C}=1)=1$
and the function is non-decreasing.

If $p_{\mathbf 1}(r)$ is highly peaked, then ${\cal C}(p_f)$ increases
rapidly as $p_f$ decreases, whereas $f(p_f)$ increases slowly. As
a result, $f({\cal C})$ is small over most of the range of $\cal
C$ as increases very rapidly to unity near the end of the range.
If, on the other hand, $p_{\mathbf 1}(r)$ is fairly uniform, then
both ${\cal C}(p_f)$ and $f(p_f)$ increase fairly abruptly over a
small range of $p_f$. The function $f({\cal C})$ then increases very
rapidly towards unity at small $\cal C$. In Figure \ref{fg.horvath}
we show the behaviour of several models of $p_{\mathbf 1}(r)$---
\begin{enumerate}
\item Some periodic functions, $\cos^2(k\cdot r)$, normalized to
   unity on two lattices; these have $\ipr=1.5$.
\item Random uncorrelated function values on sites, drawn from the
   uniform distribution in $[0,1]$, normalized to unity; these
   have $\ipr=1.66$.
\item Two Dirac eigenvectors obtained from the same gauge configuration
   at $2T_c$, one with $\ipr<2$ and the other with $\ipr>100$.
\end{enumerate}

The localization function $f({\cal C})$ clearly contains more
information than the single number $\ipr$, \ie, the IPR. However,
the IPR is statistically compatible with statements obtained from
the more detailed measurement of $f({\cal C})$. We demonstrate this
by the following correspondence. Choose any arbitrary value, ${\cal
C}_*$, the function value $f({\cal C}_*)$ is strongly correlated
with the IPR, as we show in Figure \ref{fg.iprequiv}.  For a wide
range of ${\cal C}_*$ we find $f({\cal C}_*) \propto 1/\ipr$.

We give an example of a question which can be easily answered through
the use of the localization function. If the eigenvector is localized,
then how does it fall off away from the peak? Exponential fall,
$\psi(r) \simeq \exp(-\alpha R)$, where $R$ is the distance from
the peak, would imply ${\cal C}\simeq1-g(R)\exp(-R^2)$ and $f\simeq
R^d$, in $d$ dimensions. Thus, exponential falloff of a localized
eigenvector would give rise to the relation $-\log(1-{\cal C})\propto
\sqrt f$, for both $\cal C$ and $f$ close to unity. In Figure
\ref{fg.expon}, we show that this is true of staggered Dirac
eigenvectors with $\ipr>40$ but those with $\ipr<2$ have completely
different behaviour.

A model for eigenvectors with small IPR is that of a function
$p_{\mathbf 1}(r)$ with random uncorrelated values. We call this
the site percolation model for the following reason. As we trace
out the level curves of this function by choosing $p_f$, we pick
sites independently with a probability given exactly by $f$. Each
site belongs to an unique cluster, defined as the collection of all
neighbouring sites on the lattice which are picked \footnote{The
assignment of sites to clusters is performed by the Hoshen-Kopelman
algorithm \cite{hk}.}. When $f$ is small, we find small localized
clusters, but above some critical value, we have percolating clusters.
Each realization of the random function is a realization of the
percolation problem for all possible probabilities.

As we fill a larger and larger fraction of the lattice, the number
of clusters, $N_c$, grows until the percolation threshold is reached,
after which the number of clusters begins to decrease. The clusters
are ramified, and, near the critical percolation probability, have
a fractal dimension related to the critical indices of the percolation
problem.  Above the critical porbability, the clusters have canonical
dimension, as a result of which the holes are filled in rapidly,
and $N_c$ decreases.

In Figure \ref{fg.cluster} we compare the average cluster size as
a function of $f$ for the site percolation problem and those
eigenvectors in QCD at $2T_c$ which have IPR greater than 2. The
fact that QCD has more clusters at larger $f$ than site percolation
implies that the percolating cluster constructed from $\ipr$ have
larger holes inside them where isolated clusters can exist.

\section{Stability of localization}

\begin{figure}[htb]\begin{center}
   \scalebox{0.65}{\includegraphics{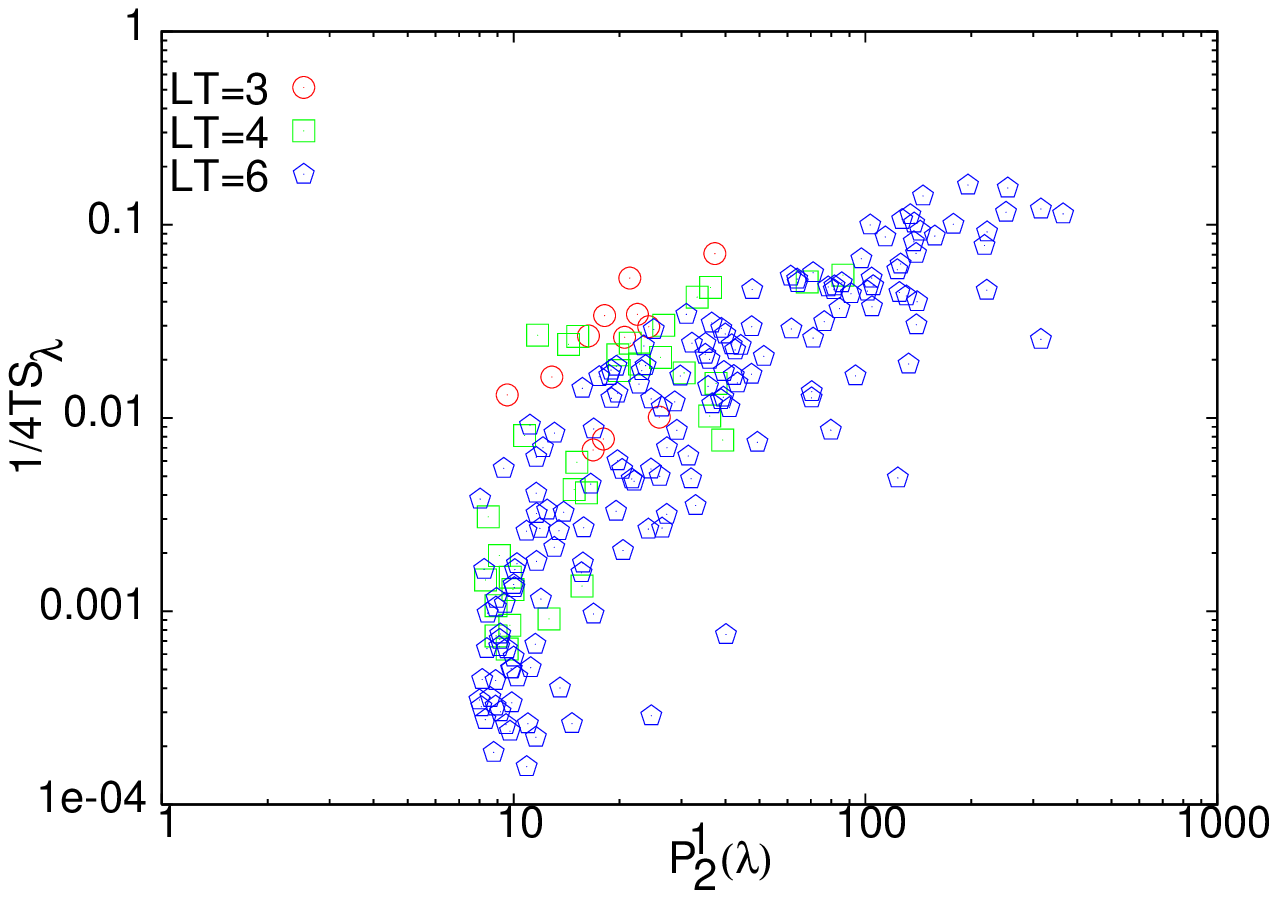}}
   \scalebox{0.65}{\includegraphics{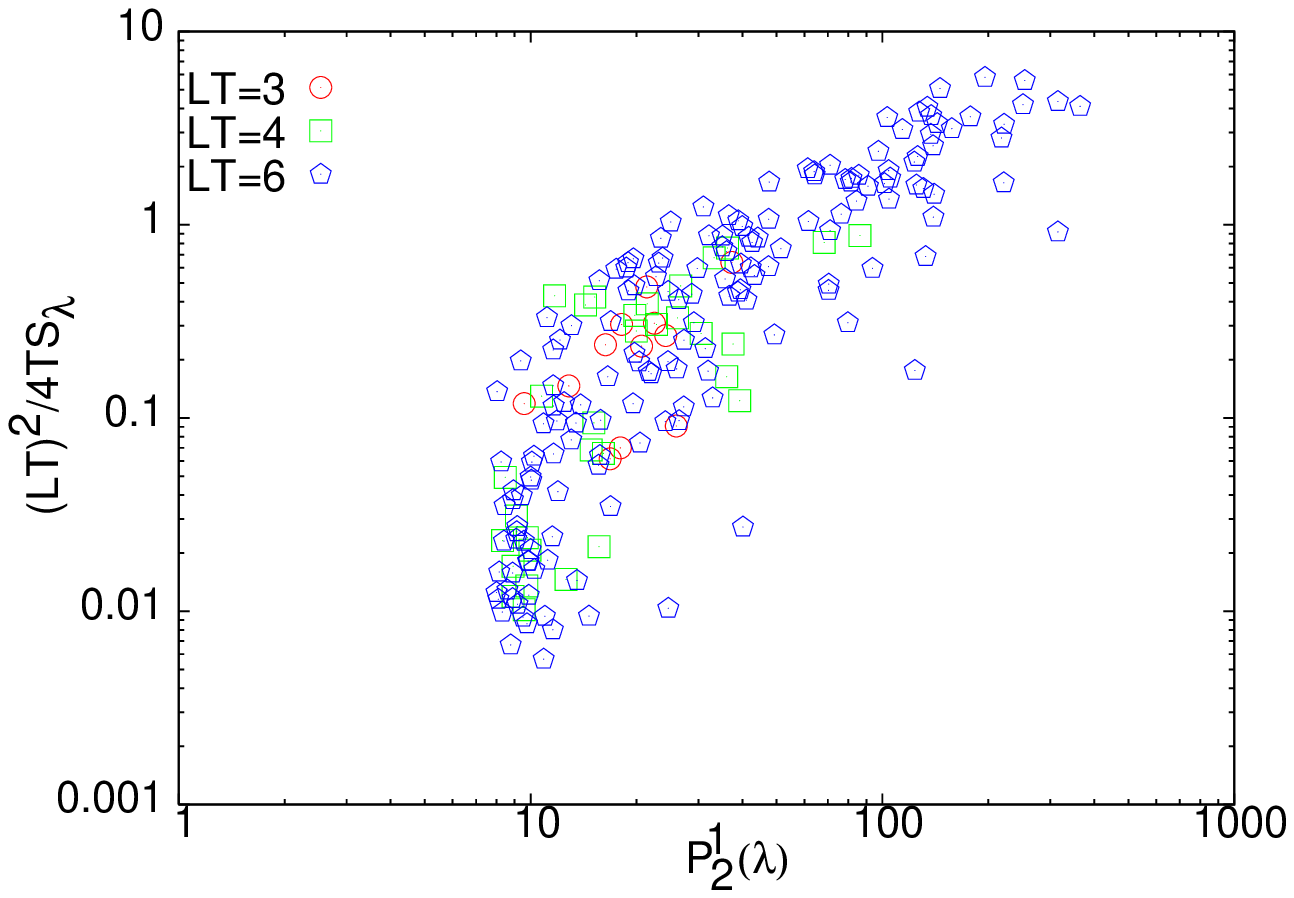}}
   \end{center}
   \caption{The stability of localization of staggered Dirac eigenvectors
    with 2 flavours of dynamical quarks in QCD at $2T_c$. The first panel
    shows that stability decreases with increasing spatial size. The second
    shows that the data supports scaling as $1/L^2$.}
\label{fg.stable}\end{figure}

\begin{figure}[htb]\begin{center}
   \scalebox{0.65}{\includegraphics{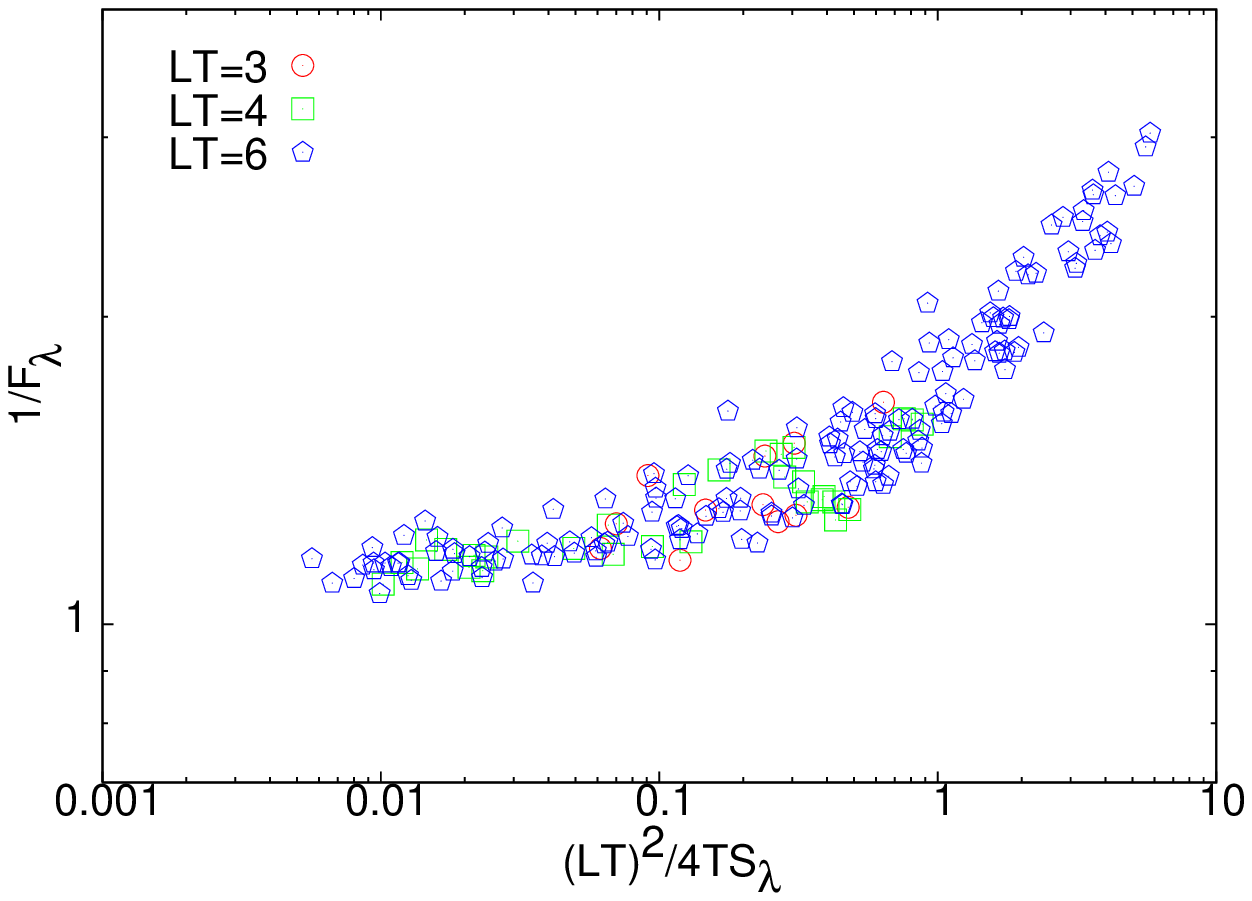}}
   \scalebox{0.65}{\includegraphics{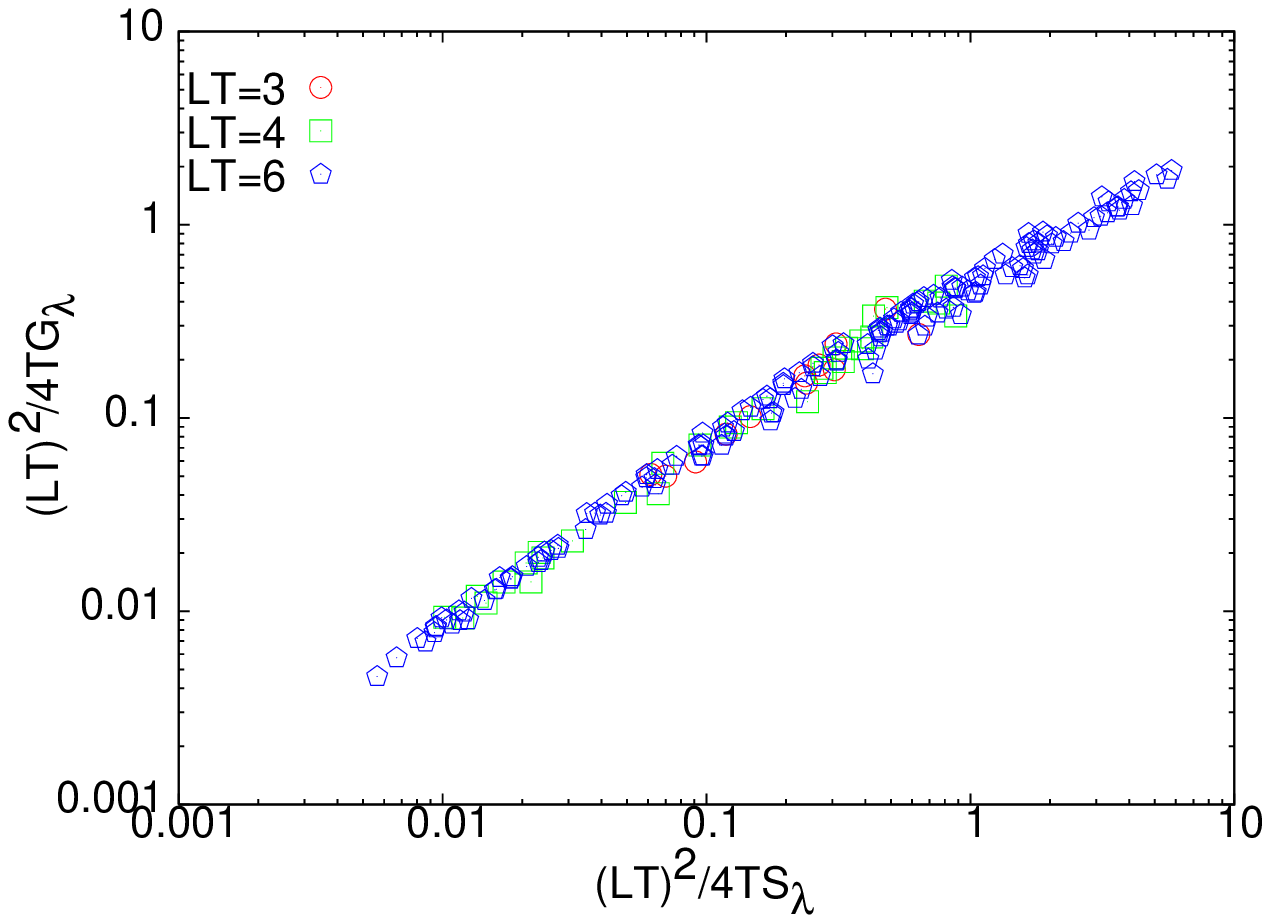}}
   \end{center}
   \caption{Correlation between the measures of stability of
    the localization of staggered Dirac eigenvectors, with 2 flavours
    of dynamical quarks in QCD at $2T_c$. Note the common scaling
    of the stability measure ${\cal S}_\lambda$ with the eigenvalue
    difference ${\cal G}_\lambda$.}
\label{fg.whystable}\end{figure}

One of the paradigms in the analysis of Dirac eigenvectors is that
of Mott localization and the existence of a mobility edge. In a
metallic crystal with random impurities, localization of electron
wavefunctions can be observed. Mott
argued that if there exist a localized and an extended state
arbitrarily close in energy, then they will mix under any small
perturbation of the Hamiltonian (induced, for example, by the
movement of one of the impurities) hence destroying localization.
He argued that, as a result, localization is robust only when
localized and extended states are separated in energy.  It is
well-known that this argument could fail if the extended states
have support in regions with holes, since the lack of overlap can
then be arranged in space rather than in energy.

The mobility edges are the band edges of localized states.  On a finite
lattice where the eigenvalue spectrum is discrete, the identification
of a mobility edge is not straightforward. As a result, it is hard to
test Mott's picture of localization directly. It is interesting to build
another measure of stability. We do this next.

Assume that the Dirac operator is perturbed by a change in the gauge
fields, $D(U+\delta U) = D(U)+\delta D$. Then, first order perturbation
theory tells us that the change in an eigenvector is
\beq
   \delta |\lambda\rangle = \sum_{\lambda'} {\mathbf C}_{\lambda\lambda'}
     |\lambda'\rangle, \qquad{\rm where}\qquad
   {\mathbf C}_{\lambda\lambda'} = \frac{\langle\lambda'|\delta D|
     \lambda\rangle}{\lambda'-\lambda}.
\label{change}\eeq
Under a random change of the gauge fields, the phase information in the
matrix element above is randomized. Hence, for a study of average properties
of the perturbations over ensembles of random changes of gauge field, it
would suffice to study $C\delta U$ instead, where
\beq
   C_{\lambda\mu} = \frac{\sum_r\sqrt{p_{\mathbf 1}(r;\lambda)
      p_{\mathbf 1}(r;\mu)}}{|\lambda-\mu|}.
\label{matrix}\eeq
This matrix can be extracted purely from the knowledge of the
eigenvalues and eigenvectors of the staggered Dirac operator. Note
also that the mixing involves both a spatial part, which is the
numerator, and a part in energy, which is the denominator. A small
mixing can be a result of either.

A perturbing field of $\delta U\simeq 1/C_{\lambda\mu}$ would change
$|\lambda\rangle$ by adding to it a significant part of $|\mu\rangle$.
As a result, the state $|\lambda\rangle$ is only as stable as the
largest value of $C_{\lambda\mu}$. The least stable eigenvector is
that for which this measure of stability is minimized. The stability
of the localization of Dirac eigenvectors in a given gauge field
configuration depends on the least stable localized eigenvector.
Hence the stability can be defined to be the quantity
\beq
   {\cal S} = {\rm min}_{\lambda\in{\mathrm loc}}{\cal S}_\lambda,
      \qquad{\rm where}\qquad
   {\cal S}_\lambda = {\rm max}_{\mu\in{\mathrm ext}} C_{\lambda\mu},
\label{stability}\eeq
such that the minimum is over states $|\lambda\rangle$ which are localized
and the maximum is over states $|\mu\rangle$ which are extended.
The inverse, $1/{\cal S}_\lambda$, for a localized state $|\lambda\rangle$,
is a measure of the minimum field strength which causes significant mixing
with an extended state.
This measure is eminently suited to a lattice where the spectrum
is discrete.  If indeed there is stable localization, then examination
of the particular element of the mixing matrix which gives $\cal
S$ can help us to identify whether localization is achieved through
Mott's mechanism and the formation of mobility edges, or through
spatial segregation of the support of localized and extended states.

A numerical implementation of eq.\ (\ref{stability}) requires
specification of which eigenvectors are localized. We use a definition
in terms of the IPR, taking all eigenvectors with $\ipr>\iprc$ are localized
and those with $\ipr<\iprc$ are deemed to be extended. When changing this
definition in the range $2\le\iprc\le 10$ we found no significant change
in the quantities reported below. The data shown in the figures are obtained
with $\iprc=8$.

In Figure \ref{fg.stable}, we show stability of the most localized states
at $2T_c$ as a function of $\ipr$. The quantity plotted is a dimensionless
measure of the minimum change in the gauge field required to mix a given
localized state with any extended state--- $1/4T{\cal S}_\lambda$. As
shown in the first panel, there is a tendency for $1/4T{\cal S}_\lambda$
at a given $\ipr$ to decrease as the lattice size increases. Scaling
the data by a power of the lattice size one finds an optimum scaling
as the point where the Fisher's linear discriminant is least able to
separate the data for different lattice sizes.  In the second panel
of the figure we exhibit the resultant scaling with the lattice size,
$1/4T{\cal S}_\lambda \propto (LT)^{-2}$ at fixed $\ipr$.  If this
scaling persists at larger lattice sizes, then it would imply that in
the thermodynamic limit an arbitrarily small change in the gauge field
can destabilize the localized eigenvalues.

The scatter in the data does not allow us to measure the scaling exponent
more precisely. One could argue that since we are examining localized states,
the factor of $p_1$ in eq.\ (\ref{matrix}) does not scale with volume. In
that case one is forced to the conclusion that the observed volume dependence
come from the energy differences in the denominator of eq.\ (\ref{matrix})
scale as $L^2$. Such a scaling is open to clear tests, and we perform this next.

Since a lattice allows only discrete eigenvalues of the Dirac operator,
the origin of localization on the lattice is not a mystery. Nevertheless,
one could try to probe the origin in more detail. In order to do this we
construct two matrices
\beq
   F_{\lambda\mu} = \sum_r\sqrt{p_{\mathbf 1}(r;\lambda)
      p_{\mathbf 1}(r;\mu)}, \qquad{\rm and}\qquad
   G_{\lambda\mu} = \frac1{|\lambda-\mu|},
\eeq
one of which, $F$, looks only at the spatial overlap, and the other,
$G$, only at the overlap in energy. Using these we can define the
notions of stability, ${\cal F}_\lambda={\rm max}_{\mu\in{\mathrm
ext}} F_{\lambda\mu}$ and ${\cal G}_\lambda={\rm max}_{\mu\in{\mathrm
ext}} G_{\lambda\mu}$. 

We found that ${\cal S}_\lambda$ is strongly correlated to both ${\cal
F}_\lambda$ and ${\cal G}_\lambda$. As expected from the earlier argument,
${\cal F}_\lambda$ shows no scaling with $L$. As a result, it requires no
scaling when plotted against the scaled quantity $L^2/{\cal S}_\lambda$.
On the other hand, whereas ${\cal G}_\lambda$ scales with the same exponent
as ${\cal S}_\lambda$. As a result, when plotted against the scaled quantity
$L^2/{\cal S}_\lambda$, one requires the scaling $L^2/{\cal G}_\lambda$ in
order for the measurements to be universal. These correlations
are shown in Figure \ref{fg.whystable}. From the figures it is clear
that the stability of the localization phenomenon seen at finite lattice
spacing is controlled by the energy level differences. The situation
seems to produce a curious version of Mott's argument. In this case
we have localized states whose spatial overlap with extended states is
finite. Thus there is no segregation of the spatial support of localized
and extended states.  Localization is seen at any finite volume, and
is realized through the formation of a mobility edge. However the gap
between the localized and extended eigenvalues seems to disappear as a power
of the lattice volume. As a result there could be no localization in the
thermodynamic limit.

One can cross check this conclusion also by computing the minimum of the
mobility gap; \ie, the difference between the maximum energy level among
the localized states and the minimum energy level between the extended
states. This mobility gap scales to zero as $1/(LT)^3$, and the scaling
is not sensitive to the choice of $P_2^*$ used to separate localized
and extended states in the range $2\le P_2^*\le10$.

\section{Conclusions}

In this paper we have examined the eigenvalues and eigenvectors of the
staggered Dirac operator evaluated on thermalized configurations obtained
in simulations of QCD with two flavours of dynamical staggered quarks
at temperatures between $0.75T_c$ and $2T_c$ with lattice spacing of
$a=1/4T$. The spectrum develops a gap as one crosses $T_c$, although in
the high temperature phase the gap remains substantially smaller than
that in free field theory. It would be interesting to study the gap
formation in the transition region to check whether this way one can
obtain additional insight on the crucial question of the order of the
phase transition.

The smallest eigenvalues have eigenvectors which are localized. We
investigated different quantities, the inverse participation ratio
(IPR) and the localization function, which measure the degree of
localization, and found good agreement between them.

We investigated the stability of localization properties of the
staggered Dirac eigenvectors with respect to changes in the gauge
field background. We showed that localization properties are not
stable as one takes the thermodynamic limit. In fact, the scaling
of the data shows that in that limit localization of staggered Dirac
eigenvectors is not expected to be of thermodynamic importance.

We developed measures of stability which distinguish between stability due
to spatial and energy separation of the eigenfunctions. QCD wth staggered
quarks seems to contain a curious reversal of Mott's argument. The
support of localized wavefunctions is not spatially separated from that
of extended wavefunctions, and this persists into the thermodynamic
limit. As a result, if localization were to be obtained, it would be
through the formation of a mobility edge. Indeed, at each volume, one
does seem to observe the formation of a mobility edge.

However, localization is spoilt by the fact that the energy denominators
can become arbitrarily small, scaling as a power of the spatial volume
in the thermodynamic limit where $L\to\infty$.  It would be interesting
to extend this work to the overlap Dirac operator, whose exact zero
modes are related to localized topological features of gauge field
configurations \cite{modes}.

This work was funded by the Indo-French Centre for the Promotion of 
Advanced Research under its project number 3104-3. Part of the
computations were carried out on the Cray X1 of the Indian Lattice
Gauge Theory Initiative (ILGTI).

\end{document}